\documentstyle[prl,aps]{revtex}

\begin{document}
\draft
\twocolumn[\hsize\textwidth\columnwidth\hsize\csname@twocolumnfalse\endcsname
\title{Introduction to the Sandpile Model}
\author{E.V.~Ivashkevich$^1$ and V.B.~Priezzhev$^2$}
\address{$^1$Dublin Institute for Advanced Studies,
10 Burlington Road, Dublin 4, Ireland\\
$^2$Laboratory of Theoretical Physics,
Joint Institute for Nuclear Research, Dubna 141980, Russia}
\date{\today}
\maketitle
\begin{abstract}
This article is based on a talk given by one of us (EVI) at the 
conference ``StatPhys-Taipei-1997''. It overviews the exact results
in the theory of the sandpile model and discusses shortly yet unsolved
problem of calculation of avalanche distribution exponents. The key
ingredients include the analogy with the critical reaction-diffusion 
system, the spanning tree representation of height configurations and
the decomposition of the avalanche process into waves of topplings. 
   
\end{abstract}
\pacs{PACS number(s): 05.40.+j}
]

\makeatletter
\global\@specialpagefalse
\def\@oddhead{\hfill DIAS-STP-97-28}
\makeatother

\section{Introduction}

The concept of Self-Organized Criticality \cite{bak-87};
one of the most fundamental concepts of modern physics of non-equilibrium 
phenomena; actually has a long history, dating back about a half of the
century, when Kolmogorov put forward his seminal theory of isotropic and
homogeneous turbulence \cite{kolmogorov-41}. The cornerstone of his very 
simple, but 
surprisingly robust dimensional analysis, was the assumption that the fluid 
being driven by a random force evolves to a stationary state where the 
velocity 
correlation functions are universal and obey power laws, provided that the 
viscosity of the fluid is small enough to ensure the existence of the 
so-called inertial range of scales. It was assumed that within this inertial 
interval the energy is being transferring from large eddies to small ones 
without any
dissipation whatsoever. Such a non-equilibrium but stationary state, with a 
constant
flux of conserved quantities, is usually called the Flux State.
 
Later on, Kardar, Parisi and Zhang \cite{kardar-86} retraced very nearly the 
same path 
on an absolutely different ground. Namely, they considered the simple type of 
growth exemplified by a vapor-deposition process, where the growth rate is 
locally determined by the flux of particles arriving ballistically at the 
surface. They found that this system evolves to the non-equilibrium stationary
state where spatial and temporal fluctuations of the growing surface obey 
power laws. 
 
After a careful analysis of all these rather phenomenological theories, 
Bak, Tang and Wiesenfeld \cite{bak-87} have squeezed the concept of 
Self-Organized Criticality out in its most purified form. 
They introduced a cellular automaton now commonly known as "sandpile" because 
of the crude analogy
between its dynamical rules and the way sand topples when building a real
sand pile. 

The formulation of this model is given in terms of integer
height variables $z_i$ at each site of the square lattice ${\cal L}$.
Particles are added randomly and the addition of a particle increases the
height at that site by one. If this height exceeds the critical value
$z_c=4$, then the site topples, on toppling its height decreases by $4$
and the heights at each of its nearest neighbors increases by $1$. 
These may become unstable in their turn and the relaxation process continues. 
This chain reaction propagates up to the moment when all sites become stable
again. One assumes the updating to be done concurrently, with all
sites updated simultaneously. 
Open boundary conditions are usually assumed, so that the toppled boundary 
site gives one particle to each of its three neighbors while one grain drops 
out of the system. 

This system also evolves stochastically into a critical state with a constant
flux of particles on which it exhibits properties similar to 
that of a second order phase transition \cite{kadanoff-89}. 
It lacks therein any characteristic 
length or time scale and obeys power-law distributions. The critical 
properties of the flux state are independent of the initial configuration of 
the system and are determined completely by the flux rate. Unlike ordinary 
critical phenomena, no fine tuning of any other control parameters is 
necessary to arrive at this state. 

\section{Mean-Field Approximation}

\subsection{Flux State}

For our derivation of the mean-field equations it will be even more convenient 
to generalize the dynamical rules of the model \cite{ivashkev-96}. 
Namely, let us suppose that at 
each site of the lattice one of the species A, B, C or D is living.
These species represent respectively four stable heights of original model.
Then, due to the external driving force, some new particles (we will call them
$\varphi$) are added randomly into the system initiating avalanches.   
Actually, $\varphi$ are the only mobile species in this model and the
avalanche process can be described in terms of their propagation through the 
correlated media of immobile species A, B, C and D. In this propagation
the $\varphi$-species mutate all others in the sites they have visited 
and topple the critical ones according to the following rules
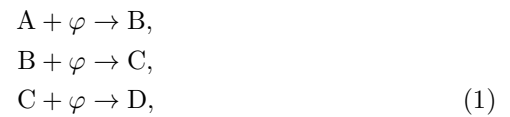
\begin{eqnarray}
{\rm A} + \varphi & \rightarrow & {\rm B}, \nonumber\\
{\rm B} + \varphi & \rightarrow & {\rm C}, \nonumber\\
{\rm C} + \varphi & \rightarrow & {\rm D}, \\
{\rm D} + \varphi & \rightarrow &
\left\{
\begin{minipage}{2.3cm}
$p_1:~ {\rm D} +  \tilde{\varphi}$\\
$p_2:~ {\rm C} + 2\tilde{\varphi}$\\
$p_3:~ {\rm B} + 3\tilde{\varphi}$\\
$p_4:~ {\rm A} + 4\tilde{\varphi}.$
\end{minipage}
\right.  \nonumber
\end{eqnarray}
Here $\varphi$ and $\tilde{\varphi}$ denote the particles obtained by the site
and the particles transferred to the neighboring sites after toppling, 
respectively.
These processes can formally be reinterpreted as an irreversible chemical
reaction which takes place at each site of the lattice or, in other words, 
as a branching process which is characterized by the so-called branching 
probabilities
\begin{equation}
{\bf p} = (p_1, p_2, p_3, p_4),~~~~~  p_1 + p_2 + p_3 + p_4 = 1.
\label{p}
\end{equation}

This generalized model also describes the critical flux state where the 
matter is conserved. It (obviously) corresponds to the total conservation of 
particles in the original formulation of the sandpile model. The latter
can be reproduced exactly if we choose ${\bf p}=(0,0,0,1)$. 

The simplest possible description of the
critical flux state can be achieved within the so-called mean-field 
approximation. 
There are a lot of different methods to calculate mean-field critical 
exponents \cite{tang-88,dhar-90a,janowsky-93,christensen-92}. 
All of them are simple 
enough and have many features in common that, when applied to our model of
chemical reactions, can be summarized as follows.

At first, we introduce the concentrations of species A, B, C and D
\begin{equation}
{\bf n} = (n_{\rm A}, n_{\rm B}, n_{\rm C}, n_{\rm D}),~~~
n_{\rm A} + n_{\rm B} + n_{\rm C} + n_{\rm D} = 1.
\end{equation}
The normalization condition comes from the constraint that we always have
only one species in each given site.

Then, following the standard prescriptions of chemical physics, we can 
write kinetic equations corresponding to this
scheme of chemical reactions
\begin{mathletters}
\begin{eqnarray}
\dot{n}_{\rm A} &=& n_\varphi~ (p_4~ n_{\rm D} - n_{\rm A}), \label{chem1}\\
\dot{n}_{\rm B} &=& n_\varphi~ (p_3~ n_{\rm D} + n_{\rm A} - n_{\rm B}), \label{chem2}\\
\dot{n}_{\rm C} &=& n_\varphi~ (p_2~ n_{\rm D} + n_{\rm B} - n_{\rm C}), \label{chem3}\\
\dot{n}_{\rm D} &=& n_\varphi~ (p_1~ n_{\rm D} + n_{\rm C} - n_{\rm D}), \label{chem4}\\
\dot{n}_\varphi &=& n_\varphi~ (\bar{p}~n_{\rm D} - 1) + \bar{p}~ \nu \nabla^2 (n_\varphi n_{\rm D}) +
\eta ({\bf r}, t) \label{chem5}
\end{eqnarray}
\end{mathletters}
where $\bar{p}=p_1+2p_2+3p_3+4p_4$ is equal to the average number of
particles $\varphi$ leaving the cell on toppling and ${\bf r}$ is the 
position vector of the site in the 2D space. The noise term $\eta({\bf r}, t)$,
being non-negative, mimics the random addition of particles to the system.
The diffusion term $\nabla^2 (n_\varphi n_{\rm D})$ describes the transfer
of particles into the neighboring cells, and the diffusion coefficient $\nu$
for the discrete Laplacian on the square lattice is equal to $1/4$.

The physical meaning of these equations is transparent. When the concentration
$n_\varphi$ of species $\varphi$ is equal to zero, all toppling processes die.
Then, due to the noise term $\eta({\bf r},t)$, the particles are added 
randomly into the system
initiating a branching process directed towards the open boundary of the
system. This process mutates species in the cells it has visited and topples
the critical ones. Finally, the system will reach the steady state where
the probability that the activity will die is on average balanced by the
probability that the activity will branch. Thus, the chain reaction maintains
this stationary state and all further avalanches cannot change the
concentrations of species A, B, C, and D. Therefore, the steady
state is characterized by the conditions that
\begin{equation}
\dot{n}_{\rm A}=\dot{n}_{\rm B}=\dot{n}_{\rm C}=\dot{n}_{\rm D}=0
\end{equation}
and Eqs.\ (\ref{chem1}-\ref{chem4}) lead to the following relations
between concentrations of species ${\bf n}$ at the stationary state and
branching probabilities ${\bf p}$
\begin{mathletters}
\begin{eqnarray}
n_{\rm A}^* &=& p_4/\bar{p},  \label{balanceA}\\
n_{\rm B}^* &=& (p_3+p_4)/\bar{p},  \label{balanceB}\\
n_{\rm C}^* &=& (p_2+p_3+p_4)/\bar{p}, \label{balanceC}\\
n_{\rm D}^* &=& (p_1+p_2+p_3+p_4)/\bar{p}=1/\bar{p}~. \label{balanceD}
\end{eqnarray}
\label{balance}
\end{mathletters}
Here, within the mean-field approximation, we obviously neglected all spatial
and temporal fluctuations of concentrations  ${\bf n}$.

\subsection{Branching Process}

The next question is how the avalanche process can be described within the 
same approximation. It can naturally be represented as the critical
branching process. To describe it we first subdivide the total number of 
sites toppled  in the avalanche into the four terms: 
$N_{\rm A}$, $N_{\rm B}$, $N_{\rm C}$ and $N_{\rm D}$. These
correspond respectively to the numbers of sites that contain species 
A, B, C and D after toppling in the avalanche. Their sum 
$N= N_{\rm A}+N_{\rm B}+N_{\rm C}+N_{\rm D}$ corresponds to the total number
of topplings in the avalanche at the moment $t$.
 
Then, we introduce the probability 
\begin{equation}
P_t(N_{\rm A},
N_{\rm B},N_{\rm C},N_{\rm D}),
\end{equation}
of having such numbers
after $t$ time steps of avalanche process and the corresponding generating 
function
\begin{eqnarray}
\lefteqn{G_t(\lambda_{\rm A},\lambda_{\rm B},
\lambda_{\rm C},\lambda_{\rm D})=}\nonumber
\\
&&~~~~~\sum_{N} P_t(N_{\rm A},
N_{\rm B},N_{\rm C},N_{\rm D}) \lambda_{\rm A}^{N_{\rm A}}
\lambda_{\rm B}^{N_{\rm B}}
\lambda_{\rm C}^{N_{\rm C}}
\lambda_{\rm D}^{N_{\rm D}}.
\label{genf}
\end{eqnarray}

Initially, at the moment $t=0$, the only nonzero element of the
probability distribution is $P_0(0,0,0,0)=1$,
which correspond to the value $G_0=1$ of the generating 
function.

On the next time step, after the toppling of the first site,
all nonzero probabilities are follows:
\begin{eqnarray}
P_1(1,0,0,0)&=&p_4 n_{\rm D},\\ 
P_1(0,1,0,0)&=&p_3 n_{\rm D}+n_{\rm A},\\ 
P_1(0,0,1,0)&=&p_2 n_{\rm D}+n_{\rm B},\\ 
P_1(0,0,0,1)&=&p_1 n_{\rm D}+n_{\rm C},
\end{eqnarray}
with corresponding generating function

\begin{eqnarray}
\lefteqn{G_{1}(\lambda_{\rm A},\lambda_{\rm B},
\lambda_{\rm C},\lambda_{\rm D})=}\\
&&
~~~~(n_{\rm D}p_4) \lambda_{\rm A}+ 
(n_{\rm D}p_3+n_{\rm A}) \lambda_{\rm B}+\\
&& 
~~~~(n_{\rm D}p_2+n_{\rm B}) \lambda_{\rm C}+ 
(n_{\rm D}p_1+n_{\rm C}) \lambda_{\rm D},
\end{eqnarray}

Now, it is easy to check directly that the generating function
obeys the following recursion relation
\begin{eqnarray}
\lefteqn{G_{t+1}=
(n_{\rm A}\lambda_{\rm B}+n_{\rm B}\lambda_{\rm C}+n_{\rm C}\lambda_{\rm D})}
\label{recursion}\\
&&~~+n_{\rm D}\left(
p_1 \lambda_{\rm D} G_t+
p_2 \lambda_{\rm C} G_t^2+
p_3 \lambda_{\rm B} G_t^3+
p_4 \lambda_{\rm A} G_t^4\right).\nonumber
\end{eqnarray}

In the limit of large avalanches we believe that 
this generating function tends to some universal function 
$G_{\infty}(\lambda_{\rm A},\lambda_{\rm B},\lambda_{\rm C},\lambda_{\rm D})$. Thus, (\ref{recursion}) becomes a closed equation for this function. 
This equation depends on two set of parameters ${\bf n}$ and ${\bf p}$. 
In the critical state, however, these are not independent. The relation 
between 
them can easily be found if we note that ``below'' the critical point, when 
all avalanches are finite, the average number of species A, B, C and D 
in the toppled sites can be simply related to the generating function 
\begin{mathletters}
\begin {eqnarray}
\langle N_{\rm A} \rangle &=& \left[\frac{\partial}{\partial\lambda_{\rm A}}
G_\infty(\lambda_{\rm A},...) \right]_{\lambda=1}=
\frac{p_4 n_{\rm D}}{1-n_{\rm D}\bar{p}}\\ 
\langle N_{\rm B} \rangle &=& \left[\frac{\partial}{\partial\lambda_{\rm B}}
G_\infty(\lambda_{\rm B},...) \right]_{\lambda=1}=
\frac{n_{\rm A}+p_3 n_{\rm D}}{1-n_{\rm D}\bar{p}}\\ 
\langle N_{\rm C} \rangle &=& \left[\frac{\partial}{\partial\lambda_{\rm C}}
G_\infty(\lambda_{\rm C},...) \right]_{\lambda=1}=
\frac{n_{\rm B}+p_2 n_{\rm D}}{1-n_{\rm D}\bar{p}}\\ 
\langle N_{\rm D} \rangle &=& \left[\frac{\partial}{\partial\lambda_{\rm D}}
G_\infty(\lambda_{\rm D},...) \right]_{\lambda=1}=
\frac{n_{\rm C}+p_1 n_{\rm D}}{1-n_{\rm D}\bar{p}}
\end{eqnarray}
\end{mathletters}
As expected, these numbers become divergent at the critical state. 
Calculating the concentrations of species A, B, C and D after the avalanche
and equating them to the concentrations before the avalanche
\begin{equation}
\frac{\langle N_{\rm A} \rangle}{
\langle N_{\rm A} \rangle+
\langle N_{\rm B} \rangle+
\langle N_{\rm C} \rangle+
\langle N_{\rm D} \rangle} = n_{\rm A}, ...
\end{equation}
we will reproduce again the relations (\ref{balance}), thus proving 
the self-consistency of this mean-field description.

It is hardly possible to solve the forth order
equation (\ref{recursion}) explicitly. We can simplify it further 
(still within the mean-field approximation) if we take into account the 
fact that 
the average number of particles leaving the site on toppling is equal to 
$\bar{p}$ and each generates on average equal sub-avalanches. 
Then, our equation for the universal generating function becomes  
\begin{equation}
G_{\infty}(\lambda)=\lambda((1-n_{\rm D})+n_{\rm D}
\left[G_\infty(\lambda)\right]^{\bar{p}}),
\label{newrecursion}
\end{equation}
where we put 
$\lambda_{\rm A}=\lambda_{\rm B}=
\lambda_{\rm C}=\lambda_{\rm D}\equiv \lambda$. 
A similar recursion relation naturally appears in any mean-field description 
of the avalanche process \cite{tang-88,dhar-90a,janowsky-93,christensen-92}. 

In the simplest case $\bar{p}=2$ the solution of 
equation 
(\ref{newrecursion}) can be written in the closed form
\begin{equation}
G_{\infty}(\lambda)=
\frac{1-\sqrt{1-4\lambda^2 n_{\rm D}(1-n_{\rm D})}}{2\lambda n_{\rm D}}.
\label{solution}
\end{equation}
which, after expanding it as a series in $\lambda$, leads to the well known 
mean-field probability distribution for the sizes of avalanches
at the critical state
\begin{equation}
P(N)\sim N^{-{3}/{2}}.
\label{mfavalanches}
\end{equation}
Actually, this asymptotic behavior does not depend on the value of $\bar{p}$
and the mean-field exponent $3/2$ is universal for any critical branching
process \cite{fisher-61}.

\section{Spanning Trees in the Sandpile Model}

\subsection{Basic Properties}

Now let us go back to the original formulation of the 
sandpile model. It has been shown by Dhar \cite{dhar-90b} that this model
is actually exactly solvable.

To derive these results it is convenient to reformulate
the dynamical rules of the sandpile model as follows. 
We consider the model on a square lattice ${\cal L}$ of $N$ sites labeled
by $1,2,...,N$. Each boundary site is connected by a bond to the additional 
site $\star$ (the {\it root}) which plays the role of the sink for the 
particles. This auxiliary site, although unphysical, will be very convenient
for all our further constructions. The $N\times N$ matrix of the  
discrete Laplacian $\Delta_{ij}$ has non-zero diagonal elements $\Delta_{ii}$
equal to the number of neighboring sites of $i$ and non-zero off-diagonal 
elements $\Delta_{ij}=-1$ for all pairs of adjacent sites $i$ and $j$. 
The addition of sand corresponds to increasing the height of the pile by
unity at a site of the lattice chosen at random (except $\star$).
If the height at any site $i$ exceeds its critical value $\Delta_{ii}$,
that site topples and all the variables $z_j,~(j=1,...,N)$ are updated 
according to the rule
\begin{equation}
z_{j}\rightarrow z_{j}-\Delta_{ij}
\end{equation}
The process stops when all the heights of the resulting configuration 
$\psi=\{z_i\}$ do not exceed their critical values.
Such a configuration is called a stable configuration. 

Yet, it is not clear whether the dynamics of the model is well defined
because during the toppling  process a configuration may occur with two
or more unstable sites. We have to make sure that the resulting stable 
configuration does not depend on the order of their topplings.
This can easily be verified if we note that after the toppling of two unstable 
sites $i$ and $j$ in an arbitrary order one gets a configuration in which 
height $z_{k}$ at any site $k$ of the lattice ${\cal L}$ decreases by 
$\Delta_{ik} + \Delta_{jk}$. This expression is invariant under
the exchange of $i$ and $j$. By a repeated use of this argument, one
obtains that after any avalanche the same final stable configuration
is produced irrespectively of the order of topplings of unstable sites
\cite{dhar-90b}. 

Another possible ambiguity is due to the fact that the updating procedure may 
in principle enter a nontrivial infinite cycle. Nevertheless, one can easily 
prove that this is impossible by noting that topplings in the interior of the 
lattice does not change the total amount of sand in the system and every
toppling on the boundary decreases this value. Here, no cycle can have
topplings at the boundary. Next, the sand on the boundary will
monotonically increase if there is any toppling one site away.
This cannot happen in the infinite cycle, thus there can be no topplings 
one site away from the edges. By induction, there can be no topplings
at an arbitrary distance from the boundary, thus, there can be no
infinite cycle
\cite{creutz-91}.

The Markovian evolution of the model implies that the set of all stable 
configurations $S$ can be divided into two subsets: recurrent $R$
and transient $S\setminus R$. By definition, the first subset includes 
those stable configurations that can be reached from all others
by sequential addition of particles. This subset is not empty because 
there is at least one such a configuration $\psi_{0}=\{z_i=\Delta_{ii}\}$ 
with all the heights equal to their critical values.
Then, the very general result of the theory of Markovian chains states that 
once the system gets into the set of reccurent configurations, it will never 
get out under the further evolution. All non-recurrent stable 
configurations are usually called transient.
 
To calculate the number of recurrent configurations 
\cite{dhar-90b}, let us consider the 
space of all possible (including unstable) configurations obtainable from
$\psi_{0}$ by addition of particles. Then, we define two such  
configurations $\{z_i\}$ and $\{z_i'\}$ as equivalent if and only if under
topplings they evolve to the same stable configuration. This means that there 
exists some integers $r_j$ such that
\begin{equation}
z_i'=z_i-\sum_{j=1}^{N} r_j \Delta_{ji},~~~{\rm for~all}~i. 
\end{equation}
Thus, if we represent configurations $\{z_i\}$ by points in a $N$-dimensional
hypercubical lattice with basis vectors $\bar{e}_i$, the set of equivalent 
points forms a super-lattice with basis vectors 
\begin{equation}
\bar{E}_i=\sum_{j=1}^N\Delta_{ij}\bar{e}_j,~~i=1~{\rm to}~N.   
\end{equation}
Since every class of equivalent configurations corresponds to some unique 
recurrent configuration, the volume of the unit cell of the super-lattice must
be equal to the number of distinct recurrent configurations. Thus one gets
\begin{equation}
{\cal N}_R = \det \Delta.  
\label{NumberReccurent}
\end{equation}

It is important to note that, although the geometrical shape of the set
$R$ in this $N$-dimensional space is quite nontrivial, the copies of $R$
can be arranged to give a simple periodic tilling of the space. 
In other words, the set $R$ of all reccurent configurations has the topology
of a $N$-dimensional torus. Therefore, the addition of a particle 
to an arbitrary site $i$ of the lattice  
can be represented by a jump of the point on this torus 
along the corresponding unit vector $\bar{e}_i$ (in the positive direction). 
Then, if one drops particles onto different sites of the lattice with equal 
probabilities, one forces the representing point on the torus to move 
randomly (but always in the positive directions of the axes) from one site  
of the torus to another. Once this random motion covers the torus
uniformly, all the reccurent configurations will appear in the Markovian 
evolution of the system with equal probability.

\subsection{Forbidden Sub-Configurations and Spanning Trees}
 
To describe in detail the set of recurrent configurations, we introduce
first the important concept of {\it forbidden sub-configuration}
\cite{majumdar-92}.
It is defined as an arbitrary subset of sites of the lattice
${\cal F}\subseteq {\cal L}$ such that all its corresponding height variables
$\{z_j\}$, $j\in {\cal F}$, satisfy the inequalities
\begin{equation}
z_{i}\leq\deg_{\cal F}(i)\equiv\sum_{j\in {\cal F},~j\neq i}(-\Delta_{ij}),~~
{\rm for~all}~ i\in {\cal F},
\label{17}
\end{equation}
where we denote by $\deg_{\cal F}(i)$ the number of bonds connecting the site 
$i$ with the other sites of the subset ${\cal F}$. 

Any stable height configuration that contains no forbidden
sub-configurations is called an allowed configuration. 
We will prove soon 
that the set of all allowed configurations $A$ is the same as
the set of all reccurent configurations $R$. 

First, let us show that the set of allowed configurations is closed
under the dynamics of the sandpile model \cite{dhar-90b}. In other words, 
we want to show 
that once the system gets into the set of allowed configurations it will
never get out. Assume the contrary. We then note that the addition of 
particles only increases heights and, hence, cannot create a forbidden 
sub-configuration. This means that there exists an allowed configuration 
$\psi$ 
such that after a single toppling it becomes the configuration $\psi'$ which 
contains a forbidden sub-configuration ${\cal F}$. Suppose the toppling 
occurs at site $i\in {\cal F}$. From the toppling rules and the definition of 
forbidden sub-configuration one gets that if ${\cal F}$ is a
forbidden sub-configuration in $\psi'$, then the set ${\cal F}\setminus i$ 
(obtained from ${\cal F}$ by deleting the site $i$) is forbidden in the 
initial configuration $\psi$.
This contradicts our assumption that $\psi$ is allowed. 
Finally, since the recurrent configuration $\psi_{0}$ is allowed 
and all other recurrent configurations are reachable from this particular one,
it follows that all recurrent configurations are allowed    
\begin{equation}
R\subseteq A~~{\rm and}~~{\cal N}_R \le {\cal N}_A. 
\label{RsubsetA}
\end{equation}

Another important notion is that of {\it toppling from the root}.
It gives a simple recursive procedure determining whether a given stable 
configuration $\psi$ is allowed. 

To organize such a toppling, we topple the auxiliary site
$\star$ as if it were an ordinary site of the lattice.
We could equivalently define it
in terms of the simultaneous dropping of particles onto each boundary site of
the lattice (i.e. one particle at each site on the 
edges and two particles onto each corner site).
If no boundary sites become unstable then the original 
configuration $\psi$ is not allowed. Using the definition one can easily 
show that in this case the lattice itself plays the role of forbidden 
sub-configuration: ${\cal F}={\cal L}$.  
More generally, if the avalanche initiated by the toppling from the boundary 
has stopped after a few topplings and some set ${\cal F}$ of lattice sites 
remains untoppled, then this set ${\cal F}$ is nothing but the forbidden 
sub-configuration associated to the initial configuration $\psi$.

In such a toppling no lattice sites can topple 
more than once \cite{ivashkev-94}. To prove this, 
let us assume that some site $i\in{\cal L}$ has toppled for the first time 
after 
its nearest neighbor $j$. Then, $i$ would topple for the second time
only after the topplings at all its neighbors, including $j$. Therefore, to
topple $i$ twice, we have first to topple the neighboring site $j$ twice. 
Using the same arguments for $j$ and for sites that toppled earlier than
$j$, we conclude that to topple $i$ more than once, we have to 
topple some boundary sites a second time. This is obviously impossible due
to the loss of particles after the first toppling of boundary sites.
 
Thus, we have proved that the initial configuration $\psi$ 
is allowed if and only if the toppling from the root generates an 
avalanche process under which each site of the lattice topples only once and, 
hence, the height configuration remains unchanged. If the initial 
configuration is not allowed, then some untoppled sites will survive and 
will form a forbidden sub-configuration.

It is possible to visualize the avalanche process initiated at the root. 
To this end let us consider the dynamical process step by step.
Initially only some of the sites at the boundary can become unstable. 
They will 
topple at the moment $t=1$ and transfer particles to neighbors that can
become unstable in their turn. 
Similarly, all sites unstable at the moment $t$ topple simultaneously and 
produce new unstable sites that will topple at the next time step $t+1$.  
Consider an arbitrary site $i$. Let $t$ be the time step at
which this site becomes unstable and $t+1$ the moment at which it topples. 
This means that at least one of its nearest neighbors toppled at the time 
step $t$.
Let $\xi$ be the number of such neighbors that toppled at the moment $t$. 
The height $z_{i}$ should obey the following
inequalities
\begin{equation}
\Delta_{ii}-\xi < z_{i} \leq \Delta_{ii},
\end{equation}
because otherwise it could not topple at the moment $t+1$ as was assumed. 
Now, if $\xi=1$, we simply mark in red the only bond connecting the site $i$ 
with the neighboring site that toppled at the earlier moment $t$.
In the other case, when $\xi > 1$, we select from
these $\xi$ possibilities only one bond to mark in red, dependent on 
the height $z_{i}$ at site $i$.
To this end we create an order of preferences by enumerating all bonds
incident to the site $i$ in an arbitrary but fixed manner. 
We then choose from the $\xi$ candidates the bond that
occupies the $(z_i+\xi-\Delta_{ii})$-th position in this list of preferences. 
For example; if the northward bond of the site $i$ 
has been allocated the number 1; the eastward 2; 
the southward 3; and the westward 4 and if the two neighboring sites 
of $i$ at the north and at the south topple at the time step $t$, then if  
$z_{i}=4$, we mark the southward bond (numbered by 3) red as it is 
the greater of the two in the list of preferences.
This algorithm makes it possible to avoid any ambiguity in the
choice of red bonds and the construction of a unique graphical representation
of a given allowed configuration.

As all sites of the allowed  configuration must topple, the
graph ${\cal T}_\star$ formed by the red bonds will cover all sites of the 
lattice.
Such a graph is usually called the {\it spanning graph}.
The fact that each site of the lattice has toppled only once implies that
the spanning graph ${\cal T}_\star$ contains no loops. Such a graph is called 
a {\it spanning tree}. A spanning tree having one site 
(the {\it root} $\star$) 
distinguished from all others is called a {\it rooted} spanning tree.
Since there is only one path from any site of the rooted spanning tree
to the root, we may uniquely orient this path so that all of its red bonds 
will be supplied with arrows in the direction of the root \cite{harary-90}.

It is not difficult to check that, given a rooted spanning tree on the 
lattice, 
one can easily reconstruct the height configuration using the same list of 
preferences as above.

Thus, starting with the definition of forbidden
sub-configurations, we have proved the one-to-one correspondence between
allowed sandpile configurations on the lattice and rooted spanning
trees on the same lattice \cite{majumdar-92}. 

\subsection{Correlation Functions and the Continuous Limit}

The spanning tree is actually a strongly
correlated object. This explains why the absolutely
uncorrelated heights of an arbitrary initial configuration 
become correlated in the recurrent state during the course of relaxation.
The tree analogy provides a useful representation for the
determination of the statistical properties of the sandpile model in
the critical state. The tool used to enumerate rooted spanning trees is 
given by the famous 

{\it  Kirchhoff's theorem}\cite{harary-90}.
If to any bond of the lattice ${\cal L}$, whose adjacent sites $i$ and $j$ are
different from the root, the weight $x_{ij}$ is ascribed, then the determinant
of the matrix
\begin{equation}
    \Delta_{ij}(x)=
        \left\{
\begin{minipage}{5.5cm}
$\sum_{k} x_{ik}$~, ~~{\rm if}~$i=j$ \\
$ -x_{ij}$~, ~~~~{\rm if}~$|i-j|=1$ \\
$0$~, ~~~~~~{\rm otherwise}.
\end{minipage}
        \right. 
    \label{kirchhoff}
\end{equation}
is a partition function of the rooted spanning trees on the lattice.

The fact that we treat here the weights $x_{ij}$ and $x_{ji}$ as different
variables implies that we consider the arrow outgoing from site $i$ and 
directed towards $j$ as different from its opposite. 
When $x_{ij}=1$ for all adjacent sites $i$ and $j$ this matrix coincides with
the discrete Laplacian and its determinant gives the total number of rooted
spanning trees. Hence the number of distinct allowed configurations
is given by
\begin{equation}
{\cal N}_A=\det \Delta.
\label{NumberAllowed}
\end{equation}
The equivalence between the set $A$ of all allowed configurations  
and the set $R$ of all reccurent configurations now follows immediately 
from Eqs.(\ref{NumberReccurent},\ref{RsubsetA},\ref{NumberAllowed}). 

Using the Kirchhoff's theorem one can determine, in principle, 
all of the correlations between 
different branches of spanning trees. Although the method of such a 
calculation is far from being novel, it is still worth recalling the
principal ideas.

Any modification of the weights of a finite number of lattice
bonds is called a local defect of the lattice. For example deleting the
bonds or inserting additional ones can be considered as a proper local defect.
The difference between a discrete Laplacian of the new lattice $\Delta'$ and
that of the initial one $\Delta$ is referred to as the defect matrix $\delta$.
The locality condition simply implies that only a finite number of the rows and
columns of the defect matrix $\delta$ have non-zero elements.
It follows that the the ratio of the number of spanning trees on the new 
lattice ${\cal L}'$
to the number of all trees on the original one ${\cal L}$ is given by an 
easily calculable determinant
\begin{equation}
    {\rm Prob}(\delta)=\frac{\det\Delta'}{\det\Delta}
     =\det({\bf 1}+G\delta)~~ ,
    \label{ProbDelta}
\end{equation}
where ${\bf 1}$ is the unit matrix and $G=\Delta^{-1}$ is the lattice Green's 
function.
On the planar square lattice the Green's function only depends on the distance 
between the sites $i$ and $j$ and  has the following integral representation
\cite{spitzer-64}
\begin{equation}
    G(n,m)-G(0,0)=
    \frac{1}{8\pi^2}\int\limits_0^{2\pi}
    \frac{\cos n\alpha~\cos m\beta - 1}{2 - \cos\alpha - \cos\beta}
    ~d\alpha d\beta.
 \label{GreenIntegral}
\end{equation}
Here $n$ and $m$ are the numbers of columns and rows between the 
sites $i$ and $j$ respectively.
This integral can be calculated explicitly with the results
\begin{equation}
G(0,1)-G(0,0)=G(1,0)-G(0,0)=-\frac{1}{4},
\end{equation}
\begin{equation}
G(n,n)-G(0,0)=-\frac{1}{\pi}\left(1+\frac{1}{3}+\frac{1}{5}+\cdots+
\frac{1}{2n-1}\right).
\end{equation}
Using these values and the discrete Laplace equation as a recursion relation,
one can find all other elements of the matrix $G_{ij}$.

For example, let us find the probability of having the height $z_i=1$
at some site $i$ deep within the lattice \cite{majumdar-91}. 
One can notice that
decreasing the height at the site $i$ by 1 --- so that its height becomes 
equal to $0$ ---  one ends up the 
forbidden sub-configuration that consists of only the site $i$ itself.
In the language of spanning trees this corresponds to those 
trees that cover site $i$ by a leaf bond (by deleting the bond 
site $i$ becomes disconnected from the rest of the tree). Due to 
the equivalence of the $4$ positions of the leaf bond, the unit height 
probability we are interested in can be expressed as
\begin{equation}
{\rm Prob}(z_i=1) = \frac{1}{4}\frac{{\cal N}'}{{\cal N}}=
\frac{1}{4}\frac{\det\Delta'}{\det\Delta}.
\label{UnitHeight}
\end{equation}
Here ${\cal N}'$ is the number of spanning trees on the new lattice 
${\cal L'}$ where all the bonds connecting the site $i$ with its four 
nearest neighbors are deleted and, instead, the site $i$ is directly 
connected to the root $\star$.
Labeling the nearest neighbors of the site $i$ in a clockwise fashion 
as $N$, $E$, $S$ and $W$, we can write the corresponding defect matrix 
\begin{equation} 
\delta=\bordermatrix{ 
~ & i & N & E & S & W \cr
i & -3& 1 & 1 & 1 & 1 \cr
N & 1 & -1& 0 & 0 & 0 \cr
E & 1 & 0 & -1& 0 & 0 \cr
S & 1 & 0 & 0 & -1& 0 \cr
W & 1 & 0 & 0 & 0 &-1 \cr
}
\label{Window}
\end{equation}
Direct evaluation of the determinant Eq.(\ref{ProbDelta}) then leads to the 
result
\begin{equation}
{\rm Prob}(z_i=1) = \frac{2}{\pi^2}\left(1-\frac{2}{\pi}\right)\approx
0.073~636.
\label{ProbUnitHeight}
\end{equation}
The calculation of all other height probabilities
requires more sophisticated ideas and can be found in \cite{priezzhev-94}.

Similarly, we could calculate the asymptotic probability that 
sites $i$ and $j$ (both deep within the lattice) separated by the distance 
$r$ will both have height $1$ in the reccurent state. 
It is more instructive however to employ another approach to find
the asymptotics of this two-point correlation function. 

Namely, we can reinterpret the partition function of the spanning
trees (given by Kirchhoff's theorem) as being the partition function of some 
artificial statistical system. 
To this end we define at each site $i$ of the lattice ${\cal L}$ the pair of 
anticommuting variables $\theta_i$ and $\theta_i^*$ (its conjugate). 
Then, using Berezin's definition of the integral over anticommuting variables
\cite{zinn-justin-93} 
we can rewrite the determinant of the matrix Eq.(\ref{kirchhoff}) as
\begin{equation}
Z=\det \Delta^{ij}=\int d\theta_1...d\theta_N^*
\exp\sum_{ij}\theta_i^* \Delta^{ij}\theta_j .
\label{BerezinIntegral}
\end{equation}
In the continuous limit this partition function defines a conformal field 
theory with the central charge $c=-2$ and the Hamiltonian 
(after integration by parts)
\begin{equation}
{\cal H}=\int \partial_\mu \theta^*\partial^\mu \theta~ {\rm d}^2 {\bf r}.
\end{equation}
The Green function of the field $\theta$ and its conjugate $\theta^*$ 
coincides with the asymptotic behavior of the lattice Green functions $G_{ij}$
\begin{equation}
\langle\theta^*({\bf r}_1)\theta({\bf r}_2)\rangle=G({\bf r}_1-{\bf r}_2)=
G({\bf 0})-\frac{1}{2\pi}\ln~|{\bf r}_1-{\bf r}_2|.
\label{GreenFunction}
\end{equation}
Changing the weight of a given directed bond in Eq.(\ref{BerezinIntegral}), 
say $x_{ij}$, and then again taking the continuous limit we will get the 
current operator corresponding to the fixed arrow on the spanning tree
\begin{equation}
j_\mu=\theta^* \partial_\mu \theta.
\end{equation}
Hence we can immediately determine the asymptotic behavior of the arrow-arrow 
correlation function on the spanning tree 
\begin{equation}
\langle j_\mu ({\bf r}) j_\nu ({\bf 0})\rangle=
\langle j_\mu \rangle \langle j_\nu \rangle +\frac{1}{4\pi^2}
\frac{x_\mu x_\nu}{r^4}.
\end{equation}
Similarly, using the defect matrix $\delta$ (\ref{Window}), we can define the 
local energy operator
\begin{equation}
\varepsilon=\partial_\mu\theta^* \partial^\mu \theta
\end{equation}
and calculate its two-point correlation function
\begin{equation}
\langle \varepsilon({\bf r}) \varepsilon({\bf 0})\rangle=
\langle \varepsilon \rangle ^2 - \frac{1}{4\pi^2}
\frac{2}{r^4}.
\end{equation}
This formula gives the asymptotic behavior 
of the two-point correlation function of unit heights in the sandpile
 model \cite{majumdar-91} 
(up to proper normalization of the local energy operator in the continuous
limit). Moreover, on the boundary of the lattice only this  
operator determines the correlations of all other heights 
\cite{ivashkev-94c,brankov-93}.

\subsection{Waves of topplings}

The study of avalanches requires a further extension of the spanning tree
representation. In order to make it possible we have to reorganize the 
topplings inside the avalanche into successive {\it waves of topplings}
\cite{ivashkev-94}. 
As has already been mentioned, the dynamics of the sandpile model admits an 
arbitrary order of topplings of unstable sites during an avalanche. 
We choose a particular ---  but quite natural ---  order amongst all others. 
Namely, let us drop a particle onto the critical site $i$ in an allowed 
configuration $\psi$. We topple it once and then topple all sites that become 
unstable keeping the initial site $i$ out of the second toppling. We call the 
set of sites toppled in this way {\it the first wave of  topplings},
denote it ${\cal F}_1$.
After the first wave has gone out we topple the site $i$ a
second time and continue the avalanche, never permitting this site
to topple a third time. The set of relaxed sites in the period
after the first wave is called {\it the second wave of topplings},
denote it ${\cal F}_2$. 
The process
continues producing intermediate configurations
$\psi_{1},\psi_{2},...$, until the site $i$ undergoes the maximum
number of topplings and the avalanche stops.

In complete analogy with the case of {\it toppling from the root}, the sites 
of any set ${\cal F}_{k}$ topple only once during the $k$-th wave.
The set ${\cal F}_{k}$ has no holes: otherwise the subset of sites
corresponding to the hole would form a forbidden sub-configuration
of the initial recurrent configuration $\psi$.
The compact cluster ${\cal F}_k$  of sites toppled in the $k$-th wave forms 
the forbidden sub-configuration ${\cal F}_{k}$ of the configuration $\psi_{k}$.
Indeed, the inequality (\ref{17}) holds for the height of every
site in the cluster provided that all other sites of the lattice 
remain untoppled during that wave.

Thus, the avalanche starting at $i$ is represented as a
collection of $T$ waves, $T \geq 1$, which we also denote by
${\cal F}_{k}$,~$(1 \leq k \leq T)$. All waves with numbers $1,...,T-1$
have the site $i$ strictly inside ${\cal F}_{k}$. The cluster ${\cal F}_{T}$ 
of the last wave, $T$, has the site $i$ on its boundary. 
Indeed, the avalanche can stop only if the site $i$ has at least one 
neighboring site not belonging to the last wave and therefore giving no 
contribution to the growth of the height $z_{i}$ in $\psi_{T}$.

To find the tree representation of waves \cite{ivashkev-94}, 
we consider the sandpile
model on a new lattice that consists of the original one ${\cal L}$ and the 
additional bond $(\star i)$ connecting a given site $i$ with the root $\star$.
Accordingly, we have to change one element of the matrix of the discrete 
Laplacian which determine the toppling rules
\begin{equation}
\Delta_{ii}\rightarrow \Delta_{ii}+1,
\end{equation}
Now, to construct the spanning tree corresponding to the wave of topplings,
let us consider again the toppling from the root $\star$. 
In this toppling, particles drop onto the boundary
sites of the lattice and one particle goes along the bond
connecting the root $\star$ with the origin of the avalanche $i$. 
Let us consider these topplings separately (again choosing a particular 
order of the topplings). We start by sending one particle to the site $i$. 
The avalanche starting at the site $i$ spreads over some part ${\cal F}_{1}$
of the lattice. 
The site $i$ can topple only once (as it is always 
connected to the root in this new lattice) hence ${\cal F}_{1}$ is 
the first wave of topplings.
We drop then all other particles onto the boundary sites causing another 
avalanche that should cover the rest of the lattice ${\cal L}\setminus 
{\cal F}_1$. This avalanche returns the configuration updated after the first 
wave to the initial state and can be termed the {\it inverse wave}.

We can generalize the procedure described above. Let us perform $k$ topplings
from the root on the new lattice with the additional bond ${(\star i)}$
and allow $k$ particles to pass through the bond ${(\star i)}$. 
They will initiate exactly $k$ waves ${\cal F}_{1},...,{\cal F}_{k}$ having 
$i$ as origin. If we then drop particles once onto the boundary sites, thus
causing an inverse wave, we will return the configuration of
heights to the previous one ($\psi_{k-1}$), corresponding to the $(k-1)$-th 
wave. 
This means that there is a complete commutativity between waves and inverse
waves and they can be performed in an arbitrary order.
Moreover, starting from a recurrent configuration on the new lattice,
by deleting one particle at site $i$ and initiating only inverse waves, one 
can reverse completely the evolution of the sandpile model.

We have seen in the previous sections that the toppling process 
naturally draws spanning trees on the lattice.
Applying the algorithm described above to the recurrent
configurations on our new lattice with the additional bond $(\star i)$, 
we get a new set of spanning
trees. All spanning trees now can be divided into two
sets. The first one consists of only those spanning trees that do not contain
the bond ${(\star i)}$ and, therefore, on deleting the bond
coincides with the set of {\it one-rooted} spanning trees ${\cal T}_\star$ 
on the old lattice ${\cal L}$.
The second set consists  of those trees that contain the bond ${(\star i)}$. 
On deleting this bond (thus returning to the original 
lattice ${\cal L}$) a subtree ${\cal T}_{i}$ gets disconnected. Considering 
the site $i$ as the second root of component ${\cal T}_{i}$ we
obtain a {\it two-rooted} situation, where a spanning tree on the
lattice consists of two disconnected clusters ${\cal T}_{i}$ and
${\cal T}_{\star}$.

Thus, in addition to the one-to-one correspondence between the recurrent
states and one-rooted spanning trees, we get the one-to-one
correspondence between all waves of topplings and all two-rooted
spanning trees. The avalanche is displayed now as a collection of
successive two-rooted trees.
 
\subsection{Waves and Green functions}

The graph representation of waves enable us to link the toppling
process and the lattice Green function $G_{ij}$.

{\it Theorem}\cite{ivashkev-94}. 
For an arbitrary lattice ${\cal L}$ with the root $\star$, the Green's
function is given by the ratio
\begin{equation}
G_{ij}= \frac{{\cal N}^{(ij)}}{{\cal N}},
\end{equation}
where ${\cal N}^{(ij)}$ is the number of two-rooted spanning trees having
the roots $\star$ and $i$, such that both vertices $i$ and $j$
belong to the same one-rooted cluster and ${\cal N}$ is the total number of
one-rooted spanning trees on ${\cal L}$.

As a simple consequence of the theorem we conclude 
that $G_{ij}$ is equal to the expected number
of topplings at the site $j$ during the avalanche caused
by adding a particle at site $i$ \cite{dhar-90b}. Indeed, as each wave 
corresponds to
only one toppling of its sites, then the expected number of topplings
should coincide with the expected number of waves involving the site $j$
and, hence, should be equal to $G_{ij}$.

Yet another result is necessary to derive the probability distributions
of waves.
Namely, as is known from conformal field theory arguments
\cite{cardy-84,saleur-87,coniglio-89}, the area $s$
and the perimeter $l$ of a finite cluster of two-rooted spanning tree 
(or any wave of topplings) are related to its radius $r$ as
\begin{equation}
s\sim r^2~~{\rm and}~~l\sim r^{5/4}.
\label{sl} 
\end{equation}   

To find the size distribution of general waves \cite{ivashkev-94}, 
we consider all
possible waves that belong to the avalanches starting at a
fixed point $i$ deep within the lattice. All these waves are in
in one-to-one correspondence with the recurrent configurations of
the sandpile model on the lattice with an additional bond
connecting the sites $\star$ and $i$. As all recurrent
configurations have the same probability it follows that all general waves are
also equally likely.

Let us now consider waves on the lattice initiated at site $i$ without
any reference to the particular avalanches they belong to. 
In other words, we consider a sequence of avalanches initiated at 
the same site $i$. These avalanches consist of waves. We ignore the pauses 
between the avalanches and study the properties of the collection of waves in 
general.
Using the above theorem we can estimate the probability
that the radius $r$ of the wave is not less 
than the distance between the sites $i$ and $j$ as
\begin{equation}
{\rm Prob}(r\geq |i - j|)\sim G_{ij}\sim \ln |i-j|.
\end{equation}
The corresponding probability density 
\begin{equation}
P_w(r)~dr\sim \frac{dr}{r}, 
\end{equation} 
is only normalizable provided that the size of the lattice $L$ (upper cutoff) 
and its spacing $a$ (lower cutoff) are fixed. 
Using scaling relations Eq.(\ref{sl}) and
\begin{equation}
P_w(s)~ds \sim P_w(l)~dl \sim P_w(r)~dr,
\label{PsPr}
\end{equation}
we can rewrite this 
probability distribution either in terms of the area $s$ or the perimeter $l$ 
of the waves of topplings.

We have seen above that the last wave corresponds to the rooted subtree 
having its root at the boundary. The alternative way to get its tree 
representation is to cut a bond and disconnect a branch 
from the one-rooted tree representing the recurrent configuration that 
appears  after the last wave. 

A general wave differs from the last one only in the location of the
root. The position of the root in the general wave is distributed
uniformly over the whole area, in contrast to its location at the boundary
for the last wave. Therefore, given a general wave of area $s$ and  
perimeter $l$, an additional factor $l/s\sim s^{-3/8}$ appears in the 
probability distribution of the last waves.
Hence, we get the size probability distribution of the last waves 
\cite{dhar-94}
\begin{equation}
P_l(s)~ds \sim P_w(s)~s^{-3/8}~ds \sim s^{-11/8}~ds.
\label{LastWavesDistribution}
\end{equation}

The critical exponents of the general waves are not related
directly to the exponents of the avalanche distributions because
by considering the general wave we lose information as to the
concrete avalanche to which it belongs. Fortunately, there exists
a situation where all avalanches consist of one and only one wave
and the avalanche distribution coincides with the distribution of
general waves. Namely, avalanches starting at the boundary consist 
of only one wave because the second toppling is
impossible due to loss of particles. 

Let us consider the case of an infinite wedge, with internal angle 
$\gamma$, related to the half plane by the conformal mapping
$w=z^{\gamma/\pi}$. It is known from the theory of complex variables 
that the Green function of the Laplacian in the region inside the wedge
has the form
\begin{equation}
G(z,\bar{z})\sim {\rm Im}(z^{-\pi/\gamma})
\end{equation}
where $z$ and $\bar{z}$,are the complex coordinates on the plane. 
Thus, the function $G(r)$ decays as
\begin{equation}
G(r) \sim r^{-\pi/\gamma}
\end{equation}
for all directions save the arms of the angle. This leads to the distribution
\begin{equation}
P_b(r)~dr \sim r^{-1-\pi/\gamma}~dr
\end{equation}
Again, using the relations $s\sim r^{2}$ and $P_b(s)~ds \sim P_b(r)~dr$,
we can rewrite this distribution in the form
\begin{equation}
P_b(s)~ds \sim s^{-1-\pi/2\gamma}~ds
\end{equation}
which corresponds to the critical exponent $\tau=1+\pi/2\gamma$ for the
area distribution of boundary avalanches \cite{ivashkev-94b}.

The angle $2\pi$ is of special interest. In this case avalanches start at 
the top of a cut on the plane. The geometry of the avalanches closely 
resembles the one occurring deep within the lattice. So one could expect
that the corresponding critical exponent $\tau=5/4$ should be close to
that occurring in the bulk of the lattice. 

\section{Avalanche distribution exponents}

Let $R$ denote the linear extent (diameter of the avalanche);
$S$, the number of distinct sites where at least one toppling occurs;
$V$, the total number of topplings in the avalanche.
We denote the number of waves of topplings it consists of as $T$.
All these quantities are random variables. It is usually assumed
that their probability distributions have the following power-law 
asymptotics in the thermodynamic limit
\begin{eqnarray}
P_a(R)~dR &\sim & R^{-\tau_R}~dR\\
P_a(S)~dS &\sim & S^{-\tau_S}~dS\\
P_a(V)~dV &\sim & V^{-\tau_V}~dV\\
P_a(T)~dT &\sim & T^{-\tau_T}~dT
\end{eqnarray}
The exponents $\tau_R, \tau_S, \tau_V$ and $\tau_T$ can be related to each 
other if we make the simple scaling assumptions that $R, S, V$ and $T$ 
scale as some powers of each other, i.e. if we assume that for all avalanches 
with a 
fixed value of one variable (say $R$) the other variables ($S, V$ and $T$) 
have strongly peaked conditional probability distributions \cite{majumdar-92}.

Since every avalanche cluster is nothing but a superposition of compact  
clusters of waves of topplings, we may speculate that the avalanche clusters
are also compact
\begin{equation}
S\sim  R^{2}.
\label{Area}
\end{equation}
The total number of topplings in the avalanche can then be related to 
its diameter via some new exponent $y$  
\begin{equation}
V\sim  R^{2+y},~~~{\rm where}~~~y\geq 0.
\label{Volume} 
\end{equation}
Finally, from the fact that in each wave of topplings the sites of the 
lattice topple only once, we conclude that the same exponent should appear
in the relation between the number of waves in the avalanche and its linear 
extent
\begin{equation}
T\sim  R^{y}.
\label{Time} 
\end{equation}
The probability distribution $P_a(T)$ of the number of waves in the avalanche 
should be consistent with these scaling relations. It should also
lead  to the logarithmic dependence of the average number of waves on the 
lattice size, $\langle T\rangle \sim \ln L$. The only distribution
that meets all these assumptions is
\begin{equation}
P_a(T)~dT \sim \frac{dT}{T^2}.
\end{equation}
Now, using scaling relations Eq.(\ref{Area},\ref{Volume},\ref{Time}) we 
can express the critical exponents of all other probability distributions
in terms of the only unknown scaling exponent $y$ 
\begin{eqnarray}
{\tau_R}&=&1+y,\\
{\tau_S}&=&1+y/2,\\
{\tau_V}&=&{(2+2y)}/{(2+y)},\\
{\tau_T}&=&2.
\end{eqnarray}
In the particular case $y=0$ these exponents coincide with those of waves
of topplings. In the opposite case, $y=1$, they reproduce the mean-field
critical exponents. Thus, we may expect that the probability distribution of 
avalanches should corresponds to some intermediate value $0\leq y \leq 1$.

The more detailed description of the avalanche process needs new
assumptions about the average space-time structure of the avalanche process
and its decomposition into the waves of topplings. On the basis of
numerical simulations it can be assumed that, typically,  the avalanche 
process consists of two regimes. In the first regime one witnesses a
relatively fast growth of the avalanche with each subsequent wave escaping the 
boundary of its  predecessor, while in the second regime one observes a 
relatively slow squeezing of waves towards the last wave of topplings that
touches the initial site. Assuming that in the thermodynamic limit the second 
stage completely dominates the first one can expect that the area decrement 
$\Delta s_t$ between two successive waves $s_t$
and $s_{t+1}$ scales along with their average size $s_k$ as
\begin{equation}
\Delta s_t\sim  s_t^{\alpha}.
\end{equation}
This scaling law is consistent with all
our earlier assumptions. The exponent $\alpha$ can be related to the scaling 
exponent $y$ that determine the avalanche probability distributions. 
Indeed, we have 
\begin{equation}
\Delta s_t\sim  s_t^{\alpha}\Delta t, ~~{\rm and }~~ 
\Delta r_t \sim r_t^{2\alpha-1} \Delta t,
\end{equation}
where $t$ is the number of wave ($\Delta t=1$ for two successive waves).   
If we integrate these relations up to the size of the avalanche 
$R$ and the total number of waves $T$, we reproduce Eq.(\ref{Time}) 
with the scaling exponent $y=2-2\alpha$.
The attempt to find $\alpha$ from more complicated scaling arguments 
\cite{priezzhev-96} gave $\alpha=3/4$ and, hence, $y=1/2$. 

Although these critical exponents are in 
agreement with the results of the real space renormalization group 
calculations \cite{pietronero-94,ivashkev-96}, 
it is worth mentioning again that the 
above construction is based only on some self-consistent hypothesis. 
It is not clear at the moment how they are supported by the numerical 
simulations.

An alternative hypothesis about inner structure of the avalanche process
assume the existence of the universal asymptotics of the probability
distribution of subsequent waves for a given size of the preceding wave,
$P(s_{t+1}|s_t)$ \cite{paczuski-97}. Namely, on the basis of 
numerical simulations it was suggested that 
\begin{equation}
P(s_{t+1}|s_t)\sim
        \left\{
\begin{minipage}{5.5cm}
$s_{t+1}^\beta$ , ~~{\rm if}~~$s_{t+1}\ll s_t$ \\
$s_{t+1}^\gamma$ , ~~{\rm if}~~$s_{t+1}\gg s_t$ 
\end{minipage}
        \right. 
\end{equation}
with $\beta=3/4$ and $\gamma=5/4$. The derivation of this scaling laws, 
if they exist, requires further analysis of the avalanche process. 

We would like to thank Deepak Dhar for many  
discussions and correspondence and James Harris for useful comments on the 
manuscript. 
EVI is grateful to Prof. Chin-Kun Hu for kind invitation to make this
talk at the conference ``StatPhys-Taipei-1997''.
VBP thanks Dublin Institute for Advanced Studies for kind hospitality.


\begin{thebibliography}{99}
\bibitem{bak-87} P.~Bak, C.~Tang, and K.~Wiesenfeld, 
  Phys. Rev. Lett. {\bf 59} (1987) 381.
\bibitem{kolmogorov-41} A.N.~Kolmogorov, 
  C. R. Acad. Sci. USSR {\bf 30} (1941) 301.
\bibitem{kardar-86} M.~Kardar, G.~Parisi and Y.-C.~Zhang, 
  Phys. Rev. Lett. {\bf 56},889 (1986).
\bibitem{kadanoff-89} L.P.~Kadanoff, S.R.~Nagel, L.~Wu and S.M.~Zhou,
  Phys. Rev. A {\bf 39} (1989) 6524.
\bibitem{ivashkev-96} E.V.~Ivashkevich,
  Phys. Rev. Lett. {\bf 76} (1996) 3368.
\bibitem{tang-88} C.~Tang and P.~Bak, Phys. Rev. Lett. {\bf 60} (1988) 2347.
\bibitem{dhar-90a} D.~Dhar and S.N.~Majumdar, 
  J. Phys. A {\bf 23} (1990) 433.
\bibitem{janowsky-93} S.A.~Janowsky and C.A.~Laberge, 
  J. Phys. A {\bf 26} (1993) L973.
\bibitem{christensen-92}K.~Christensen and Z.~Olami, 
  Phys. Rev. E {\bf 48} (1993) 3361.
\bibitem{zhang-89} Y.-C.~Zhang, Phys. Rev. Lett. {\bf 63} (1989) 470.  
\bibitem{fisher-61}  M.E.~Fisher and J.W.~Essam, 
  J. Math. Phys. {\bf 2} (1961) 609.
\bibitem{dhar-90b} D.~Dhar,   
  Phys. Rev. Lett. {\bf 64} (1990) 1613.
\bibitem{creutz-91} M.~Creutz, Comput. Phys. {\bf 5} (1991) 198.
\bibitem{majumdar-92} S.N.~Majumdar and D.~Dhar,   
  Physica A {\bf 185} (1992) 129.
\bibitem{ivashkev-94} E.V.~Ivashkevich, D.V.~Ktitarev and V.B.~Priezzhev,
  Physica A {\bf 209} (1994) 347.
\bibitem{fortuin-72} C.M.~Fortuin and P.W.~Kasteleyn, 
  Physics {\bf 57} (1972) 536.
\bibitem{harary-90}
  F.~Harary,{\it Graph Theory} (Reading, MA: Addison Wesley, 1990).
\bibitem{spitzer-64} F.~Spitzer, {\it Principles of Random Walk}
  (Van Nostrand, New York, 1964).
\bibitem{majumdar-91} S.N.~Majumdar and D.~Dhar,   
  J. Phys. A {\bf 24} (1991) L357.
\bibitem{priezzhev-94} V.B.~Priezzhev,   
  J. Stat. Phys. {\bf 74} (1994) 955.
\bibitem{zinn-justin-93} J.~Zinn-Justin, {\it Field Theory and
  Critical Phenomena} (Oxford, Clarendon, 1993).
\bibitem{ivashkev-94c} E.V.~Ivashkevich,
  J. Phys. A {\bf 27} (1994) 3643.
\bibitem{brankov-93} J.G.~Brankov, E.V.~Ivashkevich and V.B.~Priezzhev,
  J.Phys. I France {\bf 3} (1993) 1729.
\bibitem{cardy-84}
  J.L.~Cardy,{\it Phase Transitions and Critical Phenomena} vol {\bf 11},
  eds C Domb and J L Lebowitz (London: Academic Press, 1987)
\bibitem{saleur-87} H.~Saleur and B.~Duplantier,
  Phys. Rev. Lett. {\bf 58} (1987) 2325.
\bibitem{coniglio-89} A.~Coniglio,    
  Phys. Rev. Lett. {\bf 62} (1989) 3054.
\bibitem{dhar-94} D.~Dhar and S.S.~Manna, 
  Phys. Rev. E {\bf 49} (1994) 2684.
\bibitem{ivashkev-94b} E.V.~Ivashkevich, D.V.~Ktitarev and V.B.~Priezzhev,
  J. Phys. A  {\bf 27} (1994) L585.
\bibitem{pietronero-94} L.~Pietronero, A.~Vespignani and S.~Zapperi,
  Phys. Rev. Lett. {\bf 72} (1994) 1690.
\bibitem{priezzhev-96} V.B.~Priezzhev, D.V.~Ktitarev and E.V.~Ivashkevich,
  Phys. Rev. Lett. {\bf 76} (1996) 2093.
\bibitem{paczuski-97} M.~Paszuski and S.~Boettcher, 
  Phys. Rev. E (to appear).
\end{thebibliography}
\end{document}